\documentclass{article}
\usepackage[T1]{fontenc} % add special characters (e.g., umlaute)
\usepackage[utf8]{inputenc} % set utf-8 as default input encoding
\usepackage{ismir,amsmath,cite,url}
\usepackage{graphicx}
\usepackage{color}
\usepackage{comment}
\usepackage{booktabs}
\usepackage{enumitem}
\usepackage{tabularx}
\usepackage{multirow}
\usepackage{array}
\newcommand{\floor}[1]{\lfloor #1 \rfloor}

\newcommand{\vectorize}[1]{\overrightarrow{\mathbf{ #1 }}}
\newcolumntype{R}{>{\raggedleft\arraybackslash}X}
\newcolumntype{L}{>{\raggedright\arraybackslash}X}
\newcolumntype{C}{>{\centering\arraybackslash}X}
\DeclareMathOperator*{\argmax}{arg\,max}

\usepackage{lineno}
\title{The Jazz Transformer on the Front Line:\\Exploring the Shortcomings of AI-composed Music\\through Quantitative Measures}

\multauthor{Shih-Lun Wu$^{1,2}$ and Yi-Hsuan Yang$^{2,3}$} {
  $^1$ National Taiwan University,~
  $^2$ Taiwan AI Labs,~
  $^3$ Academia Sinica, Taiwan\\
{\tt\small b06902080@csie.ntu.edu.tw, yang@citi.sinica.edu.tw}
}
\begin{document}

\maketitle
\begin{abstract}
This paper presents the Jazz Transformer, a generative model that utilizes a neural sequence model called the Transformer-XL for modeling lead sheets of Jazz music. Moreover, the model endeavors to incorporate structural events present in the Weimar Jazz Database (WJazzD) for inducing structures in the generated music. While we are able to reduce the training loss to a low value, our listening test suggests however a clear gap between the ratings of the generated and real compositions. We therefore go one step further and conduct a series of computational analysis of the generated compositions from different perspectives. This includes analyzing the statistics of the pitch class, grooving, and chord progression, assessing the structureness of the music with the help of the fitness scape plot, and evaluating the model's understanding of Jazz music through a MIREX-like continuation prediction task. Our work presents in an analytical manner why machine-generated music to date still falls short of the artwork of humanity, and sets some goals for future work on automatic composition to further pursue.
\end{abstract}

\begin{figure}
 \centerline{
 \includegraphics[trim={0 0 0 0},clip,width=\linewidth,height=3.6cm]{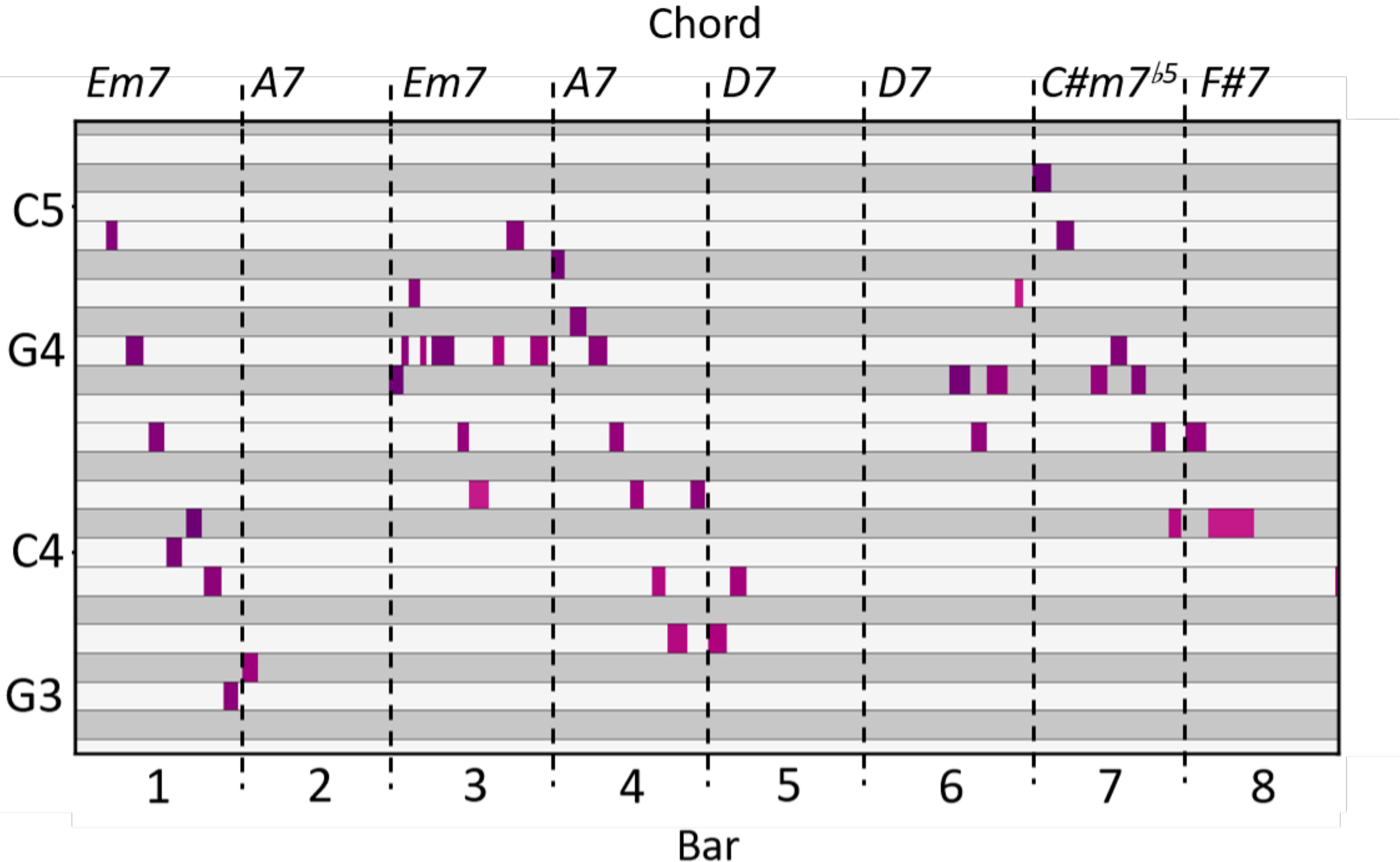}}
\caption{The first 8 bars of a piece (filename \protect\url{sample_B01.mp3} in Google Drive) composed by the Jazz Transformer, exhibiting clear rests between phrases.}
 \label{fig:pianoroll}
\end{figure}

% \begin{figure}
%  \centerline{
%  \includegraphics[trim={0 8mm 0 0},clip,width=\linewidth]{user_study_all_numbers.pdf}}
% \caption{Result of subjective study (\textbf{O:} Overall Quality, \textbf{I:} Impression, \textbf{S:} Structureness, \textbf{R:} Richness), comparing from-scratch compositions created by the proposed model with structure-related events (i.e., `Model (B)') against the real pieces from the WJazzD.}
%  \label{fig:user-study-all}
% \end{figure}

\section{Introduction}\label{sec:introduction}
Music is a heart-touching form of art that strikes a chord with people's emotions, joyful or sorrowful; intense or relieved, through the twists and turns of notes. Despite its ubiquity in our everyday lives, the composition and arrangement of music often requires substantial human effort. This is a major reason why automatic music composition is such a fascinating field of study. Over the years, researchers have sought strenuously ways for machines to generate well-formed music; such methods include meticulously designed non deep learning-based algorithms like the Markov chains \cite{anderson2013generative} and formal grammars \cite{groves2016automatic}; and, a proliferation of deep learning-based solutions in the past decade \cite{briot19book}. In this work, we exclusively study the extension and evaluation of Transformer-based models \cite{vaswani2017attention} for its claimed successes in natural language processing and music generation in recent years \cite{devlin18bert, radford19language, huang2018music, donahue2019lakhnes}.

The dataset chosen for our work is the Weimar Jazz Database (WJazzD) \cite{wjazzd,wjazzd_book}. As opposed to the commonly used piano MIDIs in recent works \cite{huang2018music, huang20remitransformer}, the choice of this dataset represents a fresh endeavor to train the Transformer on Jazz music, and grants us the unique opportunity to integrate structure-related events, precisely annotated in the WJazzD, to the model. However, such an attempt involves no short of technical challenges, including the quantization of the numerous short notes in Jazz improvisations; and, dealing with the complex chord representations used in the WJazzD. In Section \ref{sec:jazz-transformer}, we will elaborate on how these difficulties are tackled in a detailed manner.
% which are anatomized one by one in Section \ref{sec:jazz-transformer}. These include quantizing the numerous short notes in Jazz improvisations into fine timesteps of the 64th note, and dealing with the complex chord representations used in the WJazzD.

Furthermore, while recent works in Transformer-based music generation often praised the model's capabilities, like being able to compose ``compelling'' music, or generate pieces with ``expressiveness, coherence, and clear structures'' as claimed in \cite{huang2018music} and \cite{huang20remitransformer} respectively, rarely do we admit that the machine is still far behind humans, as shown in our user study (Section \ref{sec:subj-study}), and take a step back to ``face the music'', in other words, to identify what exactly goes wrong in the model's compositions.

Therefore, the goal of the paper is two-fold. First, to deploy Transformers to a new, more complex music genre, Jazz, asking the model to compose melody lines, chord progression, and structures all at once. Second, to develop a set of objective metrics (Section \ref{sec:obj-metrics}) that evaluate the generated music's pitch usages, rhythmicity, consistency in chord progression, and structureness (see Sec. \ref{subsec:struct-indic} for definition), to discover the culprits behind the model's incompetence (Section \ref{sec:expr-result-discuss}).

Figure \ref{fig:pianoroll} shows an example of a composition generated by our model, in which we may find reasonable combinations of chords and melody; and, clear rests between phrases. Audio samples can be found in a Google Drive folder,\footnote{\url{https://drive.google.com/drive/folders/1-09SoxumYPdYetsUWHIHSugK99E2tNYD?usp=sharing}} which we encourage readers to listen to. We have also open-sourced our implementation of the Jazz Transformer\footnote{\url{https://github.com/slSeanWU/jazz_transformer}} and the proposed objective metrics.\footnote{\url{https://github.com/slSeanWU/MusDr}}

\section{Related Work}\label{sec:related-work}

There has been a great body of research work on computational analysis of human performance of Jazz \cite{abesser15ISMIR,gregorio2016PhraseLevelAS,abesser17AES,weissBAM18,eEremenkoDBS18}. One prominent example is the Jazzomat %Research
Project \cite{abesser13CMMRa}, which established the WJazzD \cite{wjazzd} to study the creative processes underpinning Jazz solo improvisations \cite{wjazzd_book}. Weiss \emph{et al.} \cite{weissBAM18}, for instance, used the dataset to explore the evolution of tonal complexity of Jazz improvisations in the past century. See Sec. \ref{subsec:dataset} for more details of the dataset.

The use of Transformer-like architectures for training music composition models has drawn increasing attention recently. These works enhanced the Transformer's capability in modeling music through relative positional encoding schemes \cite{huang2018music, payne2019musenet}, cross-domain pre-training \cite{donahue2019lakhnes}, and event token design \cite{donahue2019lakhnes,huang20remitransformer}. To the best of our knowledge, this work represents the first attempt in the literature to employ Transformers to compose exclusively Jazz music.

Automatic composition of general lead sheets has been investigated lately, mostly based on recurrent neural network (RNN) models \cite{lim17ismir,boom20arxiv,liu18ismir_lbd}. As for inducing structures in the generated music, several RNN-based solutions have also been proposed \cite{medeotCKMASNW18,jhamtani19ml4md,raja20arxiv}. Since Transformers have been shown to outperform RNNs in various tasks \cite{huang2018music,chen19ismir,karita19asru}, we strive to be the forerunner in bringing them to these realms of research.

Relatively little work has been done to train a model for Jazz composition.  JazzGAN \cite{jazzgan} is a model employing a generative adversarial network (GAN) architecture for
chord-conditioned melody composition, using a dataset of only 44 lead sheets, approximately 1,700 bars. Another model presented in \cite{hung19apsipa} explores the use of recurrent variational auto-encoders for generating both the melody and chords of a lead sheet from scratch.

A number of objective metrics have been employed for measuring the performance of deep learning for music composition \cite{sturm17evaluation,chuan18aaai,musegan,yang18evaluation}. However, most of them focused on surface-level statistics only (e.g., pitch class histograms, note onset intervals, etc.). The introduction of structureness indicators and the MIREX-like metric (see Sec. \ref{subsec:struct-indic}--\ref{subsec:mirex-cont}) in this paper provide new insights into assessing music's quality at piece level, and evaluating the model's overall understanding of a certain music genre.

\section{The Jazz Transformer}\label{sec:jazz-transformer}
Transformers use self-attention modules to aggregate information from the past events when predicting the next events \cite{vaswani2017attention,dai2019transformer,katharopoulos2020transformers}. Accordingly, it is natural that we model music as a language, namely, to represent each composition by a sequence of event tokens. In this section, we will explain in detail how we break down the components of the WJazzD \cite{wjazzd_book} to construct the vocabulary of events, and how the pieces are converted into sequences that can be fed into a Transformer-like model for training.

\subsection{Dataset}\label{subsec:dataset}
The WJazzD dataset \cite{wjazzd,wjazzd_book}  comprises of 456 monophonic Jazz solos. Each solo is arranged in the lead sheet style and comes with two tracks: the melody track and the beat track. The melody track contains every note's pitch, onset time and duration (in seconds), with additional information on loudness (in decibels), phrase IDs and ``midlevel units'' (MLUs) \cite{frieler2016midlevelAO}, a structure of finer granularity than a phrase to capture the distinctive  short-time ideas in Jazz improvisations. The beat track contains the beat onsets (in seconds), chord progressions and form parts (or sections, e.g., \texttt{A1}, \texttt{B1}). The highlight of this dataset is that all the contents, including the notes, chords, metrical and structural markings, are human-annotated and cross-checked by the annotators \cite{wjazzd_book}, ensuring the data cleanliness that is often crucial for machine learning tasks. To simplify the subsequent processings, we retain only the pieces marked solely with 4/4 time signature, resulting in 431 solos. For objective analysis, we leave 5\% of the solos as the held-out validation data. See Table \ref{tab:dataset} for the statistics of the data.

\begin{table}[t]
\centering
\begin{tabularx}{\linewidth}{l R R R R}
\toprule
& \# solos & Total duration & Total \# events & Avg. \# events per solo\\
\midrule
\textbf{Train} & 409 &  11h 19m & 1,220 K & 2,983\\
\textbf{Val.} & 22 & 33m & 56 K & 2,548\\
\bottomrule
\end{tabularx}
\caption{Statistics of the dataset we compile from the WJazzD \cite{wjazzd_book}. See Section \ref{subsec:data-repr} for details of the ``events''.}\label{tab:dataset}
\end{table}

\subsection{Data Representation}\label{subsec:data-repr}
The event representation adopted here is a modified version of the ``REvamped MIDI-derived event representation'' recently proposed in \cite{huang20remitransformer}, extended to integrate the chord system and structural events of WJazzD. The resulting event encodings can be broken down into the following 4 categories: \textbf{note-related}---\textsc{Note-Velocity},~\textsc{Note-On},~\textsc{Note-Duration}; \textbf{metric-related}---\textsc{Bar},~\textsc{Position}, \textsc{Tempo-Class},~\textsc{Tempo}; \textbf{chord-related}---\textsc{Chord-Tone},~\textsc{Chord-Type},~\textsc{Chord-Slash}; and \textbf{structure-related}---\textsc{Phrase},~\textsc{MLU},~\textsc{Part},~\textsc{Repetition}.

\subsubsection{Note-related Events}\label{subsubsec:note-evs}
Each note in the melody is represented by three events, i.e., \textsc{Note-Velocity}, \textsc{Note-On}, and \textsc{Note-Duration}. 

The \textsc{Note-Velocity} event decides how hard the note should be played. We derive it according to the estimated loudness (in decibels) provided by the dataset, and quantize it into 32 bins, corresponding to MIDI velocities $[3, 7, \dots, 127]$, through 
$v = \floor{\big(80 + 3\cdot(dB - 65)\big)/4}$,
where $dB$ is the decibel value of the note, and $v$, clipped such that $v \in [1, 32]$, is the resulting \textsc{Note-Velocity($v$)} event. This mapping scheme comes in handy in the process of converting the model's compositions to MIDIs.

The \textsc{Note-On} events, ranging from $0$ to $127$, correspond directly to the MIDI numbers, indicating the note's pitch. The \textsc{Note-Duration} events represent the note's length in 64th note multiples, ranging from $1$ to $32$, obtained by taking the ratio of the note's duration (in seconds) to the duration of the beat (also in seconds) where the note situates. The reason why we use such a fine-grained quantum, while previous work mainly consider only 16th note multiples (e.g., \cite{huang2018music,huang20remitransformer}), is as follows. Most notes in WJazzD are quite short, with a significant portion being 32th and 64th notes (12.9\% and 2.7\% respectively). The quantum is chosen such that the coverage of the 32 \textsc{Note-Duration} events encompasses the most notes, which is 99.6\% with our choice of the 64th note.\footnote{All notes shorter than a 64th note are discarded and those longer than a half note are clipped.}

\subsubsection{Metric-related Events}\label{subsubsec:metric-evs}
% As demonstrated in \cite{huang20remitransformer}, the combination of the \textsc{Bar} and \textsc{Position} events represent better the progression of time than the \textsc{Time-Shift} events used in \cite{huang2018music}. And, the combination of \textsc{Tempo-Class} and \textsc{Tempo} events sets the pace the music should be played.
To model the progression of time, we use a combination of \textsc{Bar} and \textsc{Position} events; as demonstrated in \cite{huang20remitransformer}, this combination leads to clearer rhythmic structure in the generated music compared to using \textsc{Time-Shift} events introduced in \cite{huang2018music}. In addition, the pace the music should be played at is set by \textsc{Tempo-Class} and \textsc{Tempo} events.

A \textsc{Bar} event is added at the beginning of each bar, and a bar is quantized into 64 subunits, each represented by a \textsc{Position} event; for example, \textsc{Position(16)} marks the start of the 2nd beat in a bar. A \textsc{Position} event occurs whenever there is a note onset, chord change, or tempo change.
% A \textsc{Position} event appears before every note onset, and at every beat onset position (i.e. \textsc{Position(0)}, \textsc{Position(16)}, etc.) to be followed by a pair of \textsc{Tempo-Class} + \textsc{Tempo} events and possibly a chord change.
It is worth mentioning that to minimize the quantization error, a note's onset position is justified with the beat it is in through the formula:
\begin{equation}\label{eqn:note-justification}
p_n = p_b + 16 \cdot \big(t_n - t_b\big) / d_b \,,
\end{equation}
where $p_b, t_b, d_b$ are the beat's position (note that $p_b \in \{0, 16, 32, 48\}$), onset time, and duration; and $t_n$ is the note's onset time. The resulting $p_n$ is then rounded to the nearest integer to determine the note's onset position.

The \textsc{Tempo-Class} and \textsc{Tempo} events always co-occur at every beat position. The 5 \textsc{Tempo-Class} events represent the general ``feeling'' of speed (i.e. fast, or slow) with interval boundaries of $[50, 80, 110, 140, 180, 320]$ beats per minute (bpm), while the 12 \textsc{Tempo} events assigned to each tempo class in evenly-spaced steps (within the interval, e.g., $50, 52.5, 55 \text{ bpm} ...)$ determine the exact pace. The events can be derived simply by taking the reciprocal of a beat's duration (provided by WJazzD). The frequent appearance of these tempo events facilitates smooth local tempo changes common in Jazz performances.

\subsubsection{Chord-related Events}\label{subsubsec:chord-evs}
Chord progressions serve as the harmonic foundation of Jazz improvisations \cite{jazz_improvise}, hence a complex chord representation system is used in the WJazzD dataset. If we were to treat each of the 418 unique chord representations present in the WJazzD as a token, the majority of chord tokens will have very few occurrences---in fact, 287 (69\%) of them appear in less than 5 solos, making it hard for the model to learn the meaning of those chords well; plus, the process of translating chords to individual notes during the conversion to MIDIs would be extremely cumbersome.

Fortunately, thanks to the detailed clarification provided in \cite{wjazzd_book}, we are able to decompose each chord into 3 events, namely, the \textsc{Chord-Tone}, \textsc{Chord-Type}, and \textsc{Chord-Slash} events, with the help of regular expressions (regex) and some rule-based exception handling.

The 12 \textsc{Chord-Tone} events, one for each note on the chromatic scale (i.e. \texttt{C}, \texttt{C\#}, \texttt{D}, ...), determine the root note, hence the tonality, of the chord. The 47 \textsc{Chord-Type} events affect the chord's quality and emotion by the different combination of notes played relative to the root note (or, \textit{key template} as we call it); e.g., the key template of a Dominant 7th chord (\textsc{Chord-Type}(\texttt{7})) is $[0, 4, 7, 10]$. Finally, the 12 \textsc{Chord-Slash} events allow the freedom to alter the bass note to slightly tweak the chord's quality. If a chord contains no slash, its \textsc{Chord-Slash} event will share the same key as its \textsc{Chord-Tone}. For instance, the chord \texttt{C7/G}, a C Dominant 7th over G, is represented by $[\text{\textsc{Chord-Tone}(\texttt{C}), \textsc{Chord-Type}(\texttt{7}), \textsc{Chord-Slash}(\texttt{G})}]$.

Note that after our decomposition, the number of unique chord-related events is greatly reduced to 71; and, the resulting set of events is still able to represent all 418 chords in WJazzD.
It is easy to use the manually-constructed key template accompanying each \textsc{Chord-Type}, together with the \textsc{Chord-Tone} and \textsc{Chord-Slash} events to map each chord to notes during the conversion to MIDIs.

\subsubsection{Structure-related Events}\label{subsubsec:struct-evs}
For the melodies, we prepend a \textsc{Phrase} event to the notes marked as the start of a phrase. The presence of phrases may be important as it informs the model to ``take a breath'' between streams of notes. And, we retain several common types and subtypes of midlevel units (e.g., \textit{line, rhythm, lick} etc.) as \textsc{MLU} events \cite{frieler2016midlevelAO}, likewise prepended to the starting note of each MLU, hoping that the model could capture the short-term note patterns described by the MLU types. \textsc{Part} and \textsc{Repetition} events are added to each beginning and end of a form part,\footnote{For example, the entire \texttt{A1} part is represented as $[$\textsc{Part-Start(A)}, \textsc{Repetition-Start(1)}, $\dots$ other events $\dots$, \textsc{Repetition-End(1)}, \textsc{Part-End(A)}$]$.} guiding the model to generate repetitive chord progression and coherent melody lines for the parts marked with the same letter.

\subsection{Model and Training Setups}\label{subsec:model}
Due to the large number of events per solo (check Table \ref{tab:dataset}), it is hard to feed the entire pieces into a Transformer at once because of memory constraint. Therefore, we choose as the backbone sequence model the Transformer-XL \cite{dai2019transformer}, an improved variant of the Transformer which introduces recurrence to the architecture. It remedies the memory constraint and the resulting context fragmentation issue by caching the computation record of the last segment, 
% \footnote{The `segment' here is for training the sequence model, not related to music.}
and allowing the current segment to attend to the cache in the self-attention process. This allows information to flow across the otherwise separated segments, inducing better coherence in the generated music.

To evaluate the effectiveness of adding the structure-related events (cf. Section \ref{subsubsec:struct-evs}), 
%which represent a remarkable feature of WJazzD, 
we consider the following two variants in our objective analysis:
\begin{itemize}
    \item \textbf{Model (A)}: trained with no  structure-related events.
    \item \textbf{Model (B)}: trained with the complete set of events.
\end{itemize}
They both consist of 12 layers, 8 attention heads and about 41 million learnable parameters. We train them on a single NVIDIA GTX 1080-Ti GPU (with 11 GB memory) with Adam optimizer, learning rate 1e$-$4, batch size 8, segment length 512 and 100\% teacher forcing. Besides, following the Music Transformer's data augmentation setting \cite{huang2018music}, we randomly transpose each solo in the range of $-$3 to $+$3 keys in every epoch. It takes roughly a full day for the negative log-likelihood losses of the models to drop to 0.25, a level at which they are able to produce music of distinctive Jazz feeling (see Section \ref{sec:expr-result-discuss} for justifications).

\section{Subjective Study}\label{sec:subj-study}
\begin{figure}
 \centerline{
 \includegraphics[trim={0 7mm 0 0},clip,width=\linewidth]{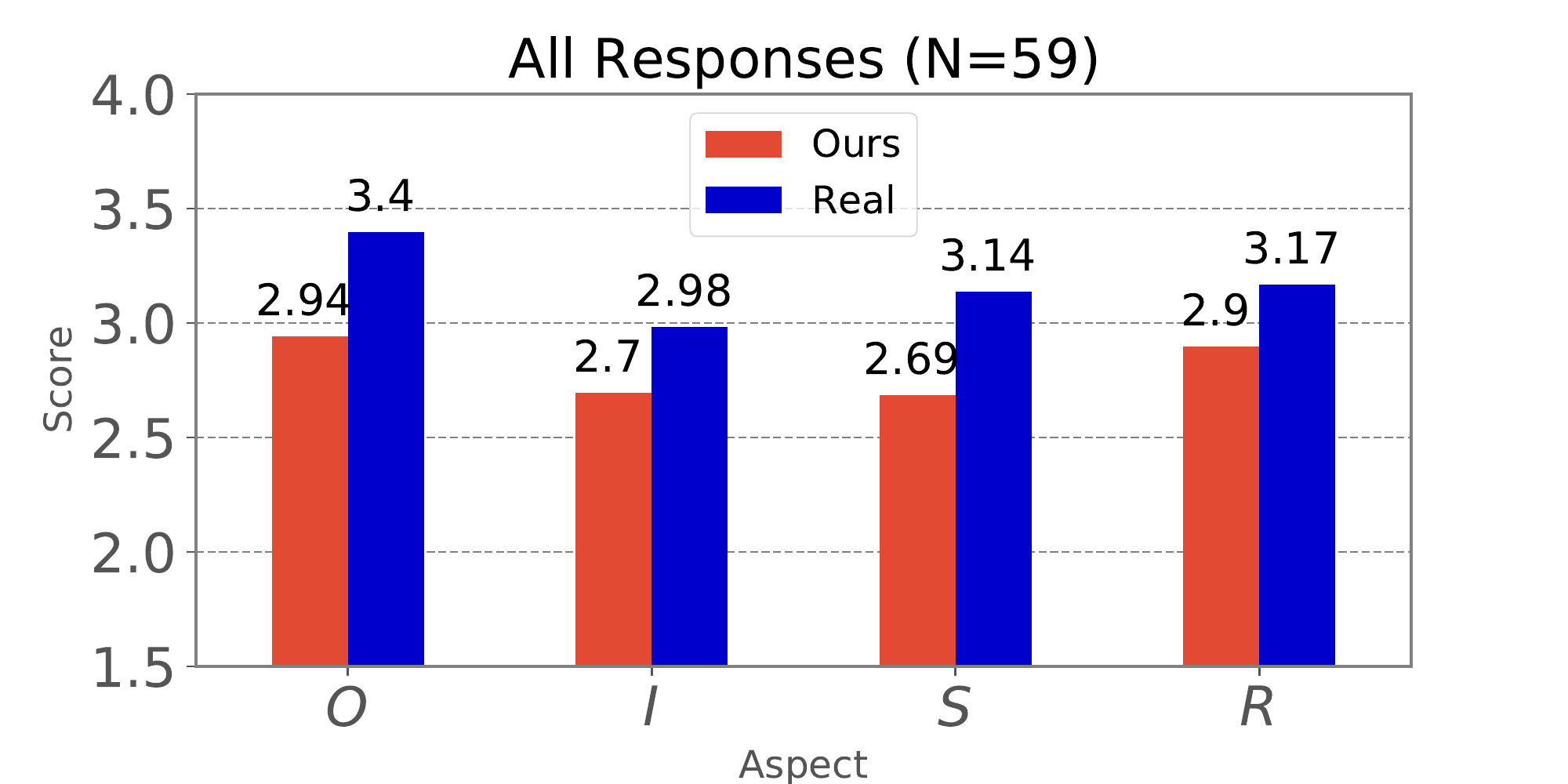}}
\caption{Result of subjective study (\textbf{O:} Overall Quality, \textbf{I:} Impression, \textbf{S:} Structureness, \textbf{R:} Richness), comparing from-scratch compositions created by the proposed model with structure-related events (i.e., `Model (B)') against the real pieces from the WJazzD. We note that the gaps in all aspects are statistically significant ($p < 0.05$).}
 \label{fig:user-study-all}
\end{figure}

To discover how users feel about the Jazz Transformer's compositions, we set up a blind listening test in which test-takers listen to four one-minute long pieces, two from the Model (B)'s compositions (at loss level 0.25), and two from real data. We do not include Model (A) here to reduce the burden on the test-takers, assuming that Model (B) is better.
We inform them that the pieces are independent of one another, and they will be asked the same set of questions after listening to each piece, namely, to rate it in a five-point Likert scale on the following aspects:
\begin{itemize}
    \item \textbf{Overall Quality (O):} Does it sound good overall?
    \item \textbf{Impression (I):} Can you remember a certain part or the melody?
    \item \textbf{Structureness (S):} Does it involve recurring music ideas, clear phrases, and coherent sections?
    \item \textbf{Richness (R):} Is the music diverse and interesting?
\end{itemize}
We distribute five test suites to our social circles and collect responses from 59 anonymized subjects, of which 27 are classified as ``pros'' for they rate their musical background (in general, not restricted to Jazz) as $4/5$ or $5/5$ (i.e., also on a five-point scale). The result shown in Figure \ref{fig:user-study-all} indicates that the Jazz Transformer receives mediocre scores and falls short of humans in every aspect, especially in \textit{overall quality} \textbf{(O)} and \textit{structureness} \textbf{(S)}.
% It is worth mentioning that the score gaps given by ``pros'' are much larger than those given by ``non-pros'' (see Table \ref{tab:user-z-test}), indicating that they are truly aware of the model's inadequacy. 
Moreover, performed \textit{one-tailed Z-tests for the difference of means} also suggests the significance of the gaps ($p < 0.05$ for all aspects), providing concrete evidence of the model's defeat.

\section{Objective Evaluation Metrics}\label{sec:obj-metrics}
The result of our subjective study poses to us an intriguing question: If the machine is still inferior to humans in creating music, then what exactly are the causes? To unravel the mystery, we develop a set of objective metrics which enables us to scrutinize the Jazz Transformer's compositions from various perspectives, and make comparisons with real data. These metrics include the analyses of event distributions, namely, the pitch class histogram, the grooving pattern, and the chord progressions; assessing the structureness with the help of the fitness scape plot; and, judging the model's performance on a discriminative task through the MIREX-like continuation prediction challenge.

\subsection{Pitch Class Histogram Entropy}\label{subsec:pitch-hist-ent}
To gain insight into the usage of different pitches, we first collect the notes appeared in a certain period (e.g., a bar) and construct the 12-dimensional pitch class histogram $\vectorize {h}$, according to the notes' pitch classes (i.e. \texttt{C}, \texttt{C\#}, ..., \texttt{A\#}, \texttt{B}), normalized by the total note count in the period such that $\sum_i{h_i} = 1$. Then, we calculate the entropy of $\vectorize{h}$:
\begin{equation}\label{eqn:pitch-entropy}
        \mathcal{H}(\vectorize{h}) = - \displaystyle \sum_{i=0}^{11} h_i \log_2 (h_i) \,.
\end{equation}
The entropy, in information theory, is a measure of ``uncertainty'' of a probability distribution \cite{shannon1948mathematical}, hence we adopt it here as a metric to help assessing the music's quality in tonality. If a piece's tonality is clear, several pitch classes should dominate the pitch histogram (e.g., the tonic and the dominant), resulting in a low-entropy $\vectorize{h}$; on the contrary, if the tonality is unstable, the usage of pitch classes is likely scattered, giving rise to an $\vectorize{h}$ with high entropy.

\subsection{Grooving Pattern Similarity}\label{subsec:groove-sim}
The grooving pattern represents the positions in a bar at which there is at least a note onset, denoted by $\vectorize g$, a 64-dimensional binary vector in our setting.\footnote{For example, if a bar contains only two note onsets, at the beginning of the 1st beat and 2nd beat respectively, then the corresponding $\vectorize g$ will have $g_0, g_{16} = 1$, and the rest dimensions $0$.} We define the similarity between a pair of grooving patterns $\vectorize g^a$, $\vectorize g^b$ as:
\begin{equation}\label{eqn:groove-sim}
    \mathcal{GS}(\vectorize g^a, \vectorize g^b) = 1 - \frac{1}{Q} \displaystyle \sum_{i=0}^{Q-1} \text{XOR}(g_i^a, g_i^b) \,,
\end{equation}
where $Q$ is the dimensionality of $\vectorize g^a$, $\vectorize g^b$, and XOR($\cdot,\cdot$) is the exclusive OR operation. Note that the value of $\mathcal{GS}(\cdot, \cdot)$ would always lie in between $0$ and $1$.

The grooving pattern similarity helps in measuring the music's rhythmicity. If a piece possesses a clear sense of rhythm, the grooving patterns between pairs of bars should be similar, thereby producing high $\mathcal{GS}$ scores; on the other hand, if the rhythm feels unsteady, the grooving patterns across bars should be erratic, resulting in low $\mathcal{GS}$ scores.

\subsection{Chord Progression Irregularity}\label{subsec:chord-irreg}
To measure the irregularity of a chord progression, we begin by introducing the term \textit{chord trigram}, which is a triple composed of 3 consecutive chords in a chord progression; for example, (\texttt{Dm7}, \texttt{G7}, \texttt{CM7}). Then, \textit{the chord progression irregularity} ($\mathcal{CPI}$) is defined as the percentage of \textit{unique chord trigrams} in the chord progression of an entire piece. Please note that 2 chord trigrams are considered different if any of their elements does not match.

It is common for Jazz compositions to make use of 8- or 12-bar-long templates of chord progressions (known as the \textit{8-}, or \textit{12-bar blues}), which themselves can be broken down into similar substructures \cite{jazz_improvise, nelson01melodic}, as the foundation of a section, and more or less ``copy-paste'' them to form the complete song with, say, \texttt{AABA} parts. Therefore, a well-composed Jazz piece should have a chord progression irregularity that is not too high.

\subsection{Structureness Indicators}\label{subsec:struct-indic}
The \textit{structureness} of music is induced by the repetitive musical content in the composition. It can involve multiple granularities, ranging from an instant musical idea to an entire section. From a psychological perspective, the appearance of repeated structures is the essence of the catchiness and the emotion-provoking nature of music \cite{levitin06brain}.

The fitness scape plot algorithm  \cite{muller11fitness, muller12scapeplot} and the associated SM Toolbox \cite{muller14smtoolbox} offer an aesthetic way of detecting and visualizing the presence of repeating structures in music. The fitness scape plot is a matrix $S_{N \times N}$,\footnote{$N$ is the number of frames sampled from the audio of a piece, the 1st axis represents the segment duration (in frames), and the 2nd axis represents the center of segment (in frames).} where $s_{ij} \in [0, 1]$ is the \textit{fitness}, namely, the degree of repeat in the piece derived from the \textit{self-similarity matrix} (SSM) \cite{foote99visualizing}, of the segment specified by $(i, j)$.

Our \textit{structureness indicator} is based on the fitness scape plot and designed to capture the most salient repeat within a certain duration interval. For brevity of the mathematical representation, we assume the sampling frame rate of $S$ is $1$ Hz (hence $N$ will be the piece's duration in seconds), and define the structureness indicator as follows:
\begin{equation}\label{eqn:struct-indic}
    \mathcal{SI}_l^u(S) = \max_{\substack{l \leq i \leq u\\ 1 \leq j \leq N}} S \,,
\end{equation}
where $l, u$\footnote{If present, otherwise $l$ defaults to $1$, and $u$ defaults to $N$.} are the lower and upper bounds of the duration interval (in seconds) one is interested in. In our experiments, we choose the structureness indicators of $\mathcal{SI}_3^8$, $\mathcal{SI}_8^{15}$, and $\mathcal{SI}_{15}$ to examine the short-, medium-, and long-term structureness respectively.

\subsection{MIREX-like Continuation Prediction Challenge}\label{subsec:mirex-cont}

Being inspired by the ``Patterns for Prediction Challenge'' held as part of the Music Information Retrieval Evaluation eXchange~(MIREX)~2019 \cite{mirex-like,janssen19ismir},
%\footnote{\url{https://www.music-ir.org/mirex/wiki/2019:Patterns_for_Prediction}} 
we developed a method to test the model's capability to predict the correct continuation given a musical prompt. The challenge is carried out as follows: First, the model is fed with the beginning 8 bars of a piece, denoted by $\vectorize{s}$; then, it is presented with a set of four 8-bar continuations $\mathcal{X} = \{\vectorize{x}^0, \vectorize{x}^1, \vectorize{x}^2, \vectorize{x}^3\}$, in which one is the true continuation, and the rest are wrong answers randomly drawn from other pieces. The way the model attempts to answer the multiple choice question is by calculating the average probability of generating the events of each continuation:
\begin{equation}\label{eqn:multi-choice-prob}
    \mathcal{P}(\vectorize{x}^i) = \frac{1}{L}\displaystyle \sum_{j=0}^{L-1} p(x_j^i \mid \Tilde{x}_{j-1}, \dots, \Tilde{x}_0; \vectorize{s}), \; i \in \{0, 1, 2, 3\} \,,
\end{equation}
where $L$ is the length of the shortest given continuation (in \# events) in $\mathcal{X}$, $x_j^i$ is the $j$-th event token in $\vectorize{x}^i$, and $\Tilde{x}_{j-1}, \dots, \Tilde{x}_0$ are the events sampled from the model's output, hence the conditional probability $p(x_j^i)$ at each timestep can be obtained straightforward. Finally, the model returns $\argmax_i \mathcal{P}(\vectorize{x}^i)$ as its answer, of which the correctness we can check.

If the model can achieve high accuracy on this continuation prediction task, we may say it possesses a good overall understanding of Jazz music, enough for it to tell right from wrong when given multiple choices.

\begin{figure}
 \centerline{
 \includegraphics[trim={0 0 0 0},clip,width=\linewidth]{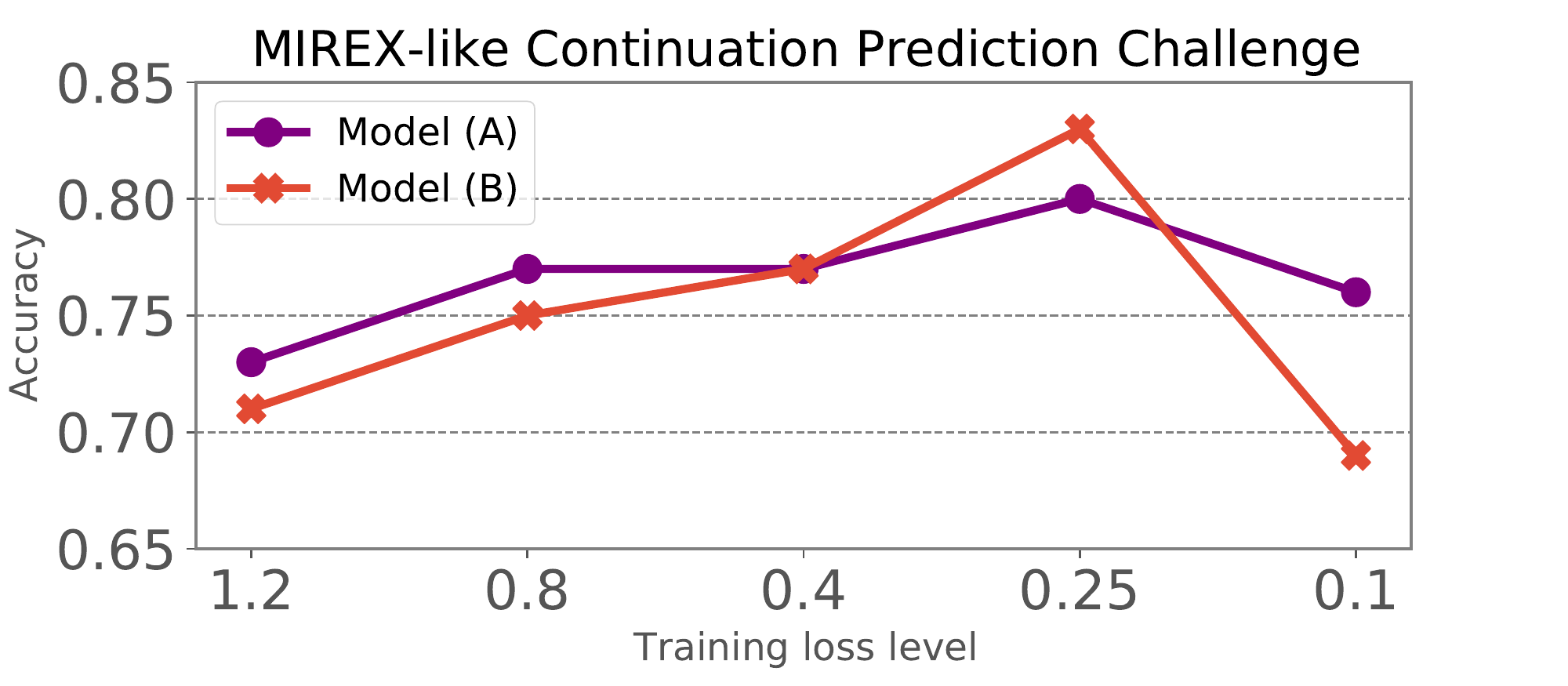}}
 \caption{Result of the MIREX-like continuation prediction challenge, each checkpoint is asked 100 questions. Notice that the accuracy of both Model (A) and (B) peaks at the loss level of 0.25, at 80\% and 83\% respectively.}
 \label{fig:multi-choice-res}
\end{figure}

\begin{table}[t]
\centering
\begin{tabularx}{\linewidth}{l | c c | c c c | C }
\toprule
& \multicolumn{2}{c |}{\textbf{Model (A)}} & \multicolumn{3}{c |}{\textbf{Model (B)}} & \textbf{Real} \\
\multicolumn{1}{r |}{\textit{loss}} & \textit{0.80} & \textit{0.25} & \textit{0.80} & \textit{0.25} & \textit{0.10} & - - \\ %\hline
\midrule
$\mathcal{H}_1$ & 2.29 & 2.45 & 2.26 & 2.20 & \textbf{2.17} & 1.94\\
$\mathcal{H}_4$ & 3.12 & 3.05 & 3.04 & \textbf{2.91} & 2.94 & 2.87\\ \hline
$\mathcal{GS}$ & \textbf{0.76} & 0.69 & 0.75 & \textbf{0.76} & \textbf{0.76} & 0.86\\ \hline
$\mathcal{CPI}$ & 81.2 & 77.6 & 79.2 & \textbf{72.6} & 75.9 & 40.4\\ \hline
$\mathcal{SI}_3^8$ & 0.18 & 0.22 & 0.25 & \textbf{0.27} & 0.26 & 0.36\\
$\mathcal{SI}_8^{15}$ & 0.15 & 0.17 & \textbf{0.18} & \textbf{0.18} & 0.17 & 0.36\\
$\mathcal{SI}_{15}$ & 0.11 & \textbf{0.14} & 0.10 & 0.12 & 0.11 & 0.35\\
\bottomrule
\end{tabularx}
\caption{Results of objective evaluations.
$\mathcal{H}_1$, $\mathcal{H}_4$ are the 1-, and 4-bar pitch class histogram entropy (see Sec. \ref{subsec:pitch-hist-ent}); $\mathcal{GS}$ is the grooving pattern similarity (Sec. \ref{subsec:groove-sim}) measured on all pairs of bars within a piece; $\mathcal{CPI}$ is the chord progression irregularity (in \%; Sec. \ref{subsec:chord-irreg}); finally, $\mathcal{SI}_3^8$, $\mathcal{SI}_8^{15}$, and $\mathcal{SI}_{15}$ are the short-, medium-, and long-term structureness indicators (Sec.  \ref{subsec:struct-indic}).
\textbf{Bold} texts indicate the model checkpoint performing the closest to real data, which is considered to be the best.
It is observed that Model (B) (i.e., the model trained with structure-related events) with a loss of 0.25 outperforms its counterparts at other loss levels and Model (A) on most of the metrics.
Moreover, consistent with the result of the MIREX-like challenge (Fig. \ref{fig:multi-choice-res}), the performance of Model (B) plunges when the loss goes too low (0.1 in this case).}\label{tab:obj-res}
\end{table}

\section{Experiment Results and Discussions}
\label{sec:expr-result-discuss}
We begin with the evaluation %exciting
on the MIREX-like challenge (Section \ref{subsec:mirex-cont}). We pick 5 checkpoints of both Model (A) and Model (B) at different training loss levels to ask each of them 100 multiple choice questions (the prompt and continuation choices of each question are randomly drawn from the held-out validation data). The result shown in Figure \ref{fig:multi-choice-res} indicates that, similarly for both models, the accuracy steadily goes up as the training loss decreases, peaks at the loss level of 0.25, and drops afterwards. This shows that the models are gradually gaining knowledge about Jazz music along the training process until a certain point, where they potentially start to overfit.

% Following the MIREX-like challenge, we pick 2 checkpoints of both Models (A) and (B), one at a relatively high loss level of 0.8; the other at the ``sweet-spot'' loss level of 0.25, for the  objective evaluations described in Sections \ref{subsec:pitch-hist-ent}--\ref{subsec:struct-indic}. In the experiments, 50  32-bar-long from-scratch compositions from each checkpointed model are compared against the 409 pieces in the training dataset.

Following the MIREX-like challenge, we pick several checkpoints of both Models (A) and (B) for objective evaluations described in Sections \ref{subsec:pitch-hist-ent}--\ref{subsec:struct-indic}. The chosen checkpoints are at loss levels 0.8, 0.25, and 0.1 (for Model (B) only, since in the MIREX-like challenge (Fig. \ref{fig:multi-choice-res}), its accuracy drastically drops when the loss reduces from 0.25 to 0.1). In the experiments, 50 32-bar-long from-scratch compositions from each checkpointed model are compared against the 409 pieces in the training dataset.

From the results (Table \ref{tab:obj-res}), we can summarize the model's deficiencies as follows: 1) the \textit{erraticity} of the generated musical events; and, 2) the \textit{absence} of medium- and long-term repetitive structures. Comparing with the real data, the first argument can be justified by the higher $\mathcal{H}_1$ and $\mathcal{H}_4$, manifesting the unstable usage of pitches at local scale; and, the lower $\mathcal{GS}$ and higher $\mathcal{CPI}$ of the entire pieces, marking the lack of consistency in rhythm and harmony from a global point of view; meanwhile, the second argument can be explained directly by the significantly lower values of structureness indicators $\mathcal{SI}_8^{15}$ and $\mathcal{SI}_{15}$, suggesting that while the model might be able to repeat some short fragments of music, creating structures of a longer time span is still beyond its capability.

Much to our delight, the introduction of structure-related events seems to be functional to some extent, noticeable from the numbers that Model (B) at 0.25 loss level is for most of the time the closest competitor to humans, with a substantial lead on the metrics focusing on shorter timespans (i.e., $\mathcal{H}_1$, $\mathcal{H}_4$, and $\mathcal{SI}_3^8$) when placed in comparison with Model (A). This suggests that the use of \textsc{Phrase} and \textsc{MLU} events provides some assistance to the model in modeling music. Furthermore, resonating with the accuracy trend in the MIREX-like challenge, the performance worsens when the loss is reduced to an overly low level.

\begin{figure}
 \centerline{
 \includegraphics[trim={0 0 0 5}, clip,width=\linewidth]{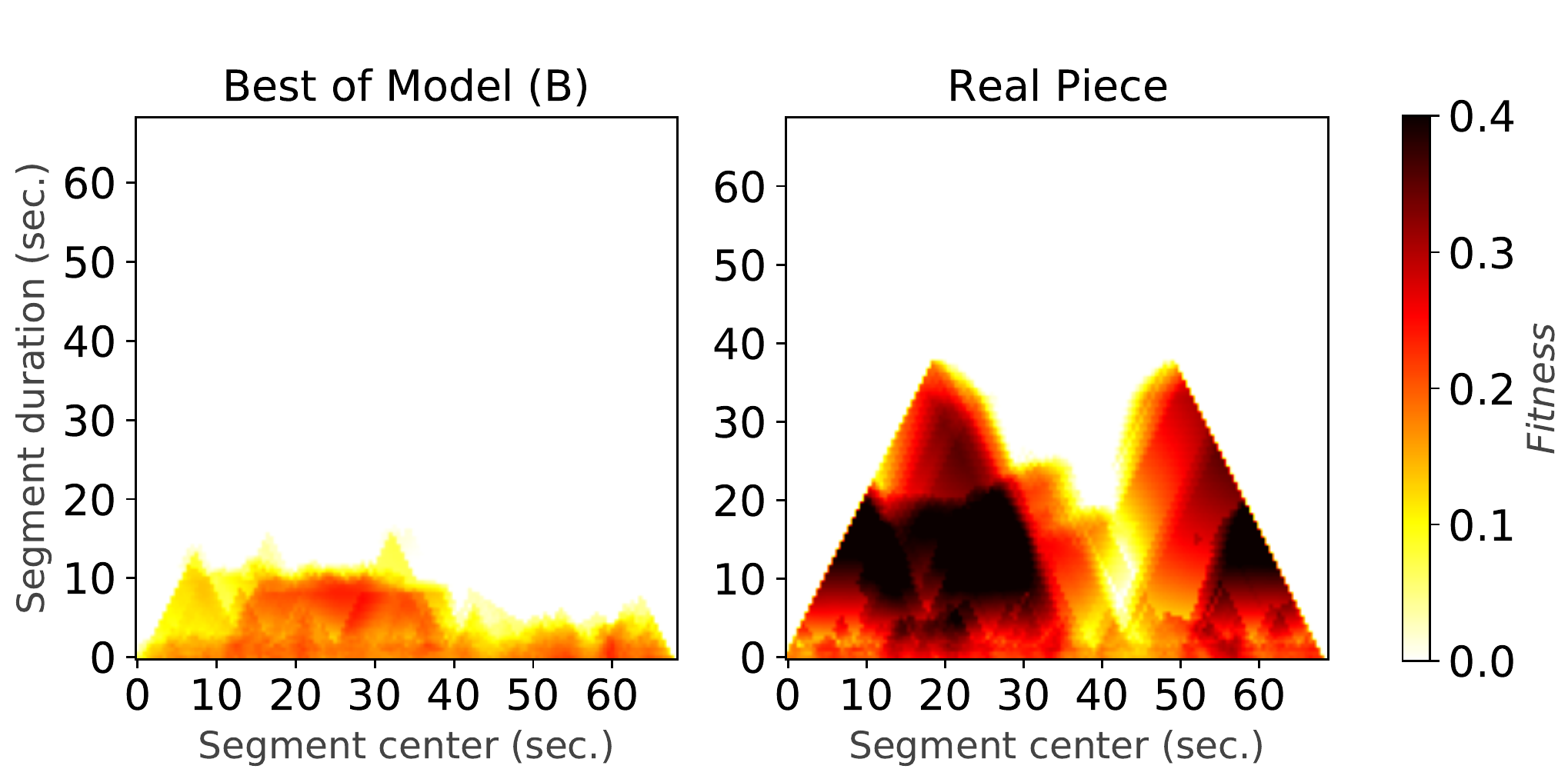}}
 \caption{The fitness scape plots of Model (B)'s best composition (according to the \textit{structureness} \textbf{(S)} score in our subjective study, see Sec. \ref{sec:subj-study}) versus a human composition in the WJazzD. Note that the piece by Model (B) contains almost no signs of repetition longer than 10 seconds, while the real piece's repetitive structures extend well into the 20--30 seconds range.}
 \label{fig:scapeplot-cmp}
\end{figure}

To visualize the deficiency in structureness of the model's compositions, we choose the piece which scores the highest, 3.14, in the \textit{structureness} \textbf{(S)} aspect in our subjective study; and, a human composition of the same duration, receiving 3.54 in the aspect \textbf{S}, for a head-to-head comparison of their fitness scape plots. The rivalry (see Figure \ref{fig:scapeplot-cmp}) reveals the stark contrast between their fitness values across all timescales. In the model's work, all traces of repetitive structures disappear at the timescale of 10 seconds; whereas in the human composition, not only do the fitness values stay high in longer timespans, but a clear sense of section is also present, as manifested by the 2 large, dark ``triangles'' in its scape plot.

% Concerning the limitations of our work, one part is that our objective metrics are yet to consider the coherence between the melody lines and their accompanying chords, which is a key element of Jazz and other types of music alike; however, designing such metric requires a thorough understanding of the theories of harmony and counterpoint, which is beyond the authors' scope of knowledge. Furthermore, although our objective evaluations partially explain the model's incompetence in generating music of good \textit{overall quality} \textbf{(O)} and \textit{structureness} \textbf{(S)} as shown in the subjective study (refer to Section \ref{sec:subj-study}), larger-scale studies are yet to be done to find out the correlations between those quantitative metrics and the aspects of subjective evaluation on the quality of music. Lastly, apart from the WJazzD, datasets featuring such complete and clean markings of musical structures are scarce, imposing restrictions on our method of treating them as events to feed into a Transformer-like architecture for training.

\section{Conclusion and Future Work}\label{sec:conclusion}
In this paper, we have presented the Jazz Transformer, whose incorporation of structure-related events has been shown useful here in enhancing the quality of machine-generated music. Moreover, we have proposed a series of objective metrics that shed light on the shortcomings of machine-composed pieces, including the erratic usage of pitch classes, inconsistent grooving pattern and chord progression; and, the absence of repetitive structures. These metrics not only show that the Transformer is in fact not that good a music composer, but also serve as effective quantitative measures for future efforts in automatic music composition to assess their models' performance, which by now still relies heavily on human evaluation.

% In the future, we plan to continue to work on inducing structures in machine-composed music; such endeavors may not stay on revamping events that fit into Transformers as done, but involve a complete redesign of the Transformer architecture, enabling it to read the structural information directly computable from data, say, the fitness scape plot, to grasp the blueprint of a piece before composing music in finer scales.

In the future, we plan to carry out larger-scale studies to explicate the correlations between those quantitative metrics and the aspects of subjective evaluation; and, to continue working on inducing structures in machine-composed music; such endeavors may not stay on revamping events that fit into Transformers as done, but involve a complete redesign of the Transformer architecture, enabling it to read the structural information directly computable from data, say, the fitness scape plot, to grasp the blueprint of a piece before composing music at finer scales.

\section{Acknowledgements}\label{sec:ack}
The authors would like to express the utmost gratitude to \textbf{the Jazzomat Research Project} (University of Music FRANZ LISZT Weimar), for compiling the WJazzD dataset and making it publicly available; \textbf{Wen-Yi Hsiao} (Taiwan AI Labs), for rendering the MIDIs to audios for subjective study; and \textbf{Yi-Jen Shih} (National Taiwan University), for the help in arranging our open-source codes.

\bibliography{ISMIRtemplate}

\end{document}